\documentclass[aps,prb,twocolumn,superscriptaddress,showpacs]{revtex4}

\usepackage{graphicx, bm, amsmath}

\newcommand{\dif}{\mathrm{d}}

\begin{document}


\title{Influence of thermal coupling on spin avalanches in Mn$_{12}$-acetate}

\author{C.\,H.\ Webster} \email{carol.webster@npl.co.uk}
\author{O.\ Kazakova} \author{\mbox{J.\,C. Gallop}} \author{P.\,W.\ Josephs-Franks}
\affiliation{National Physical Laboratory, Hampton Road, Teddington,
Middlesex, TW11 0LW, United Kingdom}

\author{A.\ Hern\'andez-M\'inguez}
\affiliation{Departament de F\'isica Fonamental, Universitat de
Barcelona, Avenguda Diagonal 647, Barcelona 08028, Spain}

\author{A.\,Ya.\ Tzalenchuk}
\affiliation{National Physical Laboratory, Hampton Road, Teddington,
Middlesex, TW11 0LW, United Kingdom}





\date{\today}

\begin{abstract}
The effect of thermal coupling on spin avalanches in
Mn$_{12}$-acetate has been probed using a single crystal assembly.
Time-resolved, synchronized measurements of magnetization and
temperature are reported.  Unusually low avalanche trigger fields
occur when thermal coupling to the bath is weak. A temperature rise
observed at zero magnetic field is attributed to a change in
magnetostatic energy.
\end{abstract}

\pacs{75.20.-g, 75.45.+j, 75.50.Xx, 75.60.Ej}


\maketitle


Over the past decade there has been much interest in single molecule
magnets due to their macroscopic quantum behaviour.
Mn$_{12}$O$_{12}$(CH$_3$COO)$_{16}$(H$_2$O)$_4$, abbreviated to
Mn$_{12}$-acetate or Mn$_{12}$, is one of the most studied single
molecule magnets to date.\cite{Christou-2000}  Strong uniaxial
anisotropy and a large spin ($S = 10$) enable macroscopic quantum
tunneling (MQT), in which the spin can reverse its direction by
tunneling through the anisotropy barrier. In bulk crystals of
Mn$_{12}$ such tunneling events can be observed as steps in the
magnetization curve, which occur at regular intervals in the
external magnetic field $B_n = nB_0$, where $n$ is an integer and
$B_0 = 0.46$ T.\cite{Friedman-1996, Thomas-1996}  Recently, however,
much attention has focussed on spin avalanches, in which the entire
ensemble of spins reverses very rapidly.\cite{Paulsen-1995} Spin
avalanches are generally accepted to be thermal phenomena, driven by
the cooperative process of spin-phonon
relaxation.\cite{Perenboom-1998, DelBarco-1999, Chiorescu-2000} They
occur only when the sample is sufficiently large that phonons
emitted by the reversing spins cannot escape to the bath.  The most
recent research has shown that avalanches propagate through the
crystal at constant velocity---a classical phenomenon dubbed
``magnetic deflagration".\cite{Suzuki-PRL-2005}

In this Brief Report we study the influence of thermal coupling to a
heat bath on spin avalanches in a single crystal assembly of
Mn$_{12}$. We report measurements of the magnetic hysteresis curve
by superconducting quantum interference device (SQUID) magnetometry
and time-resolved, synchronized measurements of the sample
temperature and magnetization, using a pickup coil. The effect of
thermal coupling on the trigger field is discussed. We also study a
temperature rise at zero magnetic field that is not associated with
avalanches. Similar temperature rises have been observed by other
researchers,\cite{Suzuki-JAP-2005, Fominaya-1997, Fominaya-1999} but
have never been satisfactorily explained.

The experiments described in this Brief Report were performed on an
assembly of single crystals of Mn$_{12}$-acetate grown according to
the procedure described by Lis.\cite{Lis-1980} Each crystal had a
length $\sim 2$ mm and a cross sectional area $\sim 0.5 \times 0.5$
mm$^2$. The long dimension is known to correspond to the $c$-axis.
Hence, we were able to glue 35 crystals together using G.\,E.\
varnish to create an assembly of mass $m = 21$ mg with the $c$-axes
aligned to within $\sim 5^{\circ}$.

We began by measuring two sets of magnetic hysteresis curves using a
commercial SQUID magnetometer.  In the first set, the sample was
strongly coupled to the bath via a constant flow of helium exchange
gas. In the second set, the thermal coupling was weakened by
enclosing the sample in an evacuated glass tube. For both sets the
sample was mounted with the $c$-axis parallel to the external
magnetic field to within $\sim 5^{\circ}$.

\begin{figure}[!ht]
\begin{center}
\includegraphics{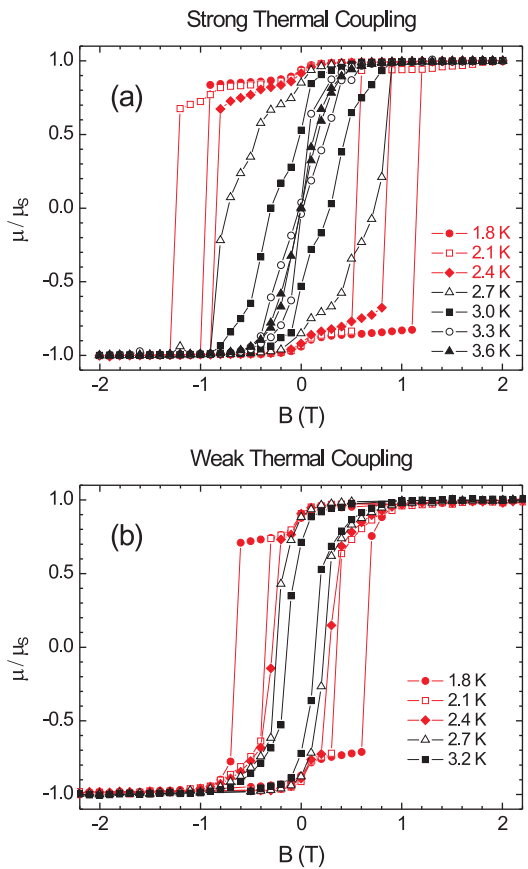}
\caption{(Color online). Magnetic hysteresis curves measured by
SQUID magnetometry. The magnetic moment is normalized to its
saturation value $\mu_{\rm S}$ at $B = 2$ T. (a) Strong thermal
coupling between sample and bath. Avalanches are observed between
1.8 and 2.4 K. Resonant MQT steps are observed between 2.7 and 3.3
K. (b) Weak thermal coupling between sample and bath. Avalanches are
observed between 1.8 and 2.4 K. Above 2.4 K the magnetization curve
is smoothly hysteretic with no discontinuities.}
\label{fig:AvalanchesWith+WithoutHe}
\end{center}
\end{figure}

For the first set of measurements we initially cooled the sample to
1.8 K with the external magnetic field set to zero.  By ramping the
field slowly to 2 T we found the saturation
moment\footnote{Throughout this Brief Report, we use the symbol
$\mu$ for the magnetic moment of the sample and $M$ for the
magnetization ($M = \mu/V$, where $V$ is the sample volume).} to be
0.95 mJ\,T$^{-1}$. We then measured a series of hysteresis curves at
different temperatures, each time sweeping from $+2$ T to $-2$ T and
back at a rate of 3 mT\,s$^{-1}$. Avalanches were consistently
observed for $T \le 2.4$ K, but the field values at which they were
triggered were not completely reproducible, giving rise to
asymmetric hysteresis curves [Fig.\
\ref{fig:AvalanchesWith+WithoutHe}(a)].  In addition, the trigger
fields did not always correspond to the resonance fields $B_{\rm
n}$.  This is in contrast to previous measurements made on
similar-sized assemblies of single crystals at similar
temperatures,\cite{Bal-Mn12-2004, Tejada-2004} but in agreement with
measurements made on individual single crystals at lower
temperatures.\cite{Suzuki-JAP-2005}  Above 2.4 K, the avalanches
disappeared and MQT steps were recovered.  Above 3.6 K, the
magnetization curve became non-hysteretic.

For the second set of measurements the thermal coupling was weakened
by enclosing the sample in an evacuated glass tube that functioned
as a shield against the flow of helium exchange gas. It was
necessary to reduce the sample volume by $\sim$ 20\% in order to fit
it inside the tube. This reduced the mass from 21 to 17 mg and the
saturation moment from 0.95 to 0.77 mJ\,T$^{-1}$. The rest of the
experiments described in this Brief Report were conducted on the
diminished sample.

Again, we measured hysteresis curves at a series of temperatures
[Fig.\ \ref{fig:AvalanchesWith+WithoutHe}(b)]. As before, spin
avalanches were observed for $T \le 2.4$ K. However, this time the
trigger fields were lower and more reproducible, leading to
symmetric hysteresis curves.  In addition, the magnetic moment
approached saturation more slowly leading to rounding of the
avalanche steps as they approached saturation.  Above 2.4 K no MQT
was observed, in contrast to the multiple steps observed with the
strongly coupled sample.  Instead, the hysteresis curve became
smoothly hysteretic. The loss of MQT in the weakly coupled sample
suggests a strong rise in sample temperature due to spin-phonon
relaxation. When thermal coupling to the bath is strong, heat is
rapidly dissipated, but when the coupling is weak the heat cannot
escape so quickly to the bath. It appears that the glass tube
provides sufficient thermal isolation that sample heating causes the
spin reversal mechanism to be dominated by thermal activation,
rather than by MQT.

There are several reports in the literature of significant rises in
the sample temperature during spin avalanches\cite{Paulsen-1995,
Tejada-2004, Bal-Mn12-2004} and MQT, \cite{Suzuki-JAP-2005,
Fominaya-1997, Fominaya-1999} with reported values ranging between 1
and 5 K. The spread is probably due to variations in sample size,
field-sweep-rate, location of the thermometer, and thermal coupling
between the sample and the bath. Here, we report time-resolved
measurements of the temperature rise during avalanches in a weakly
coupled sample. Our measurements were performed in a helium cryostat
with a superconducting magnet.  We placed the sample at the bottom
of an open-ended glass tube that functioned as a weak thermal link
to the bath.  A multiturn coil was wound around the base of the tube
to enable us to detect changes in the sample magnetization. The tube
was then placed inside an evacuated copper sample holder with a
vertical slit to minimize eddy currents, which was attached to the 1
K plate. We estimate that the $c$-axis of the crystal assembly was
aligned parallel to the magnetic field to within $\sim 10^{\circ}$.
A Cernox resistance thermometer was mounted directly on top of the
sample using high thermal conductivity grease (Apiezon N). A low
bias current of 5 nA was used to minimize self heating.

The sample was initially cooled to 1.6 K with the external magnetic
field set to zero. We then swept the field at a rate of 24
mT\,s$^{-1}$ and used a VXI data aquisition board to record
time-resolved signals from either the coil or the thermometer.
Separate measurements were subsequently synchronized by aligning
their trigger points.  The temperature and magnetization
measurements were both triggered from the thermometer signal, which
was found to be highly reproducible on successive field sweeps.  The
rate of change of the sample magnetization $\dif M/\dif t$ was
calibrated by measuring the constant voltage induced across the coil
by the external field ramp $\dif B/\dif t$. This was done at high
fields (between 2 T and 3.5 T) where the sample magnetization was
saturated.

\begin{figure}[!ht]
\begin{center}
\includegraphics{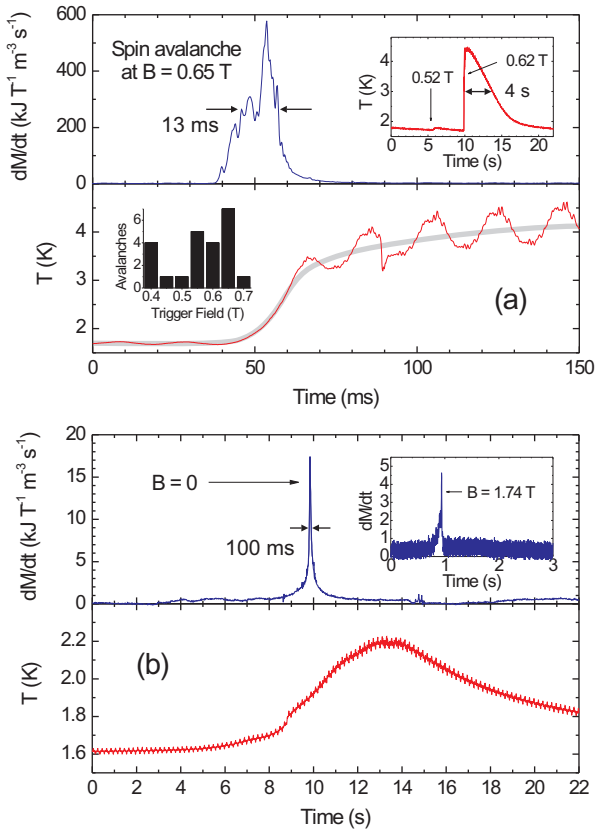}
\caption{(Color online). Time-resolved, synchronized measurements of
magnetization and temperature on a sample weakly coupled to the
bath. (a) Main plots: spin avalanche at $B = 0.65$ T.  The grey line
is a guide to the eye.  Upper inset: separate avalanche recorded
over a longer time interval. Prior to this avalanche, a small rise
in temperature occurs at $B \approx B_1$. Lower inset: distribution
of trigger fields. (b) Step at $B=0$.  This plot comprises four
separate measurements (two of temperature and two of magnetization)
which have been aligned by their trigger points.  Inset: small
change in magnetization at $B \approx B_4$.} \label{fig:Avalanche}
\end{center}
\end{figure}

Figure \ref{fig:Avalanche}(a) shows the changes in sample
temperature and magnetization observed during an avalanche. The
oscillation in the temperature signal is due to electrical pickup of
the 50 Hz line voltage. \footnote{The amplitude of the 50 Hz
oscillation gets larger as the temperature increases because the
sensor resistance decreases by an order of magnitude between 1.6 and
4.3 K. This causes the ratio of pickup voltage to sensor voltage to
decrease.} A distribution of avalanche trigger fields was observed
between 0.37 and 0.65 T, with a mean of $0.56 \pm 0.09$ T, which is
close to $B_1$. The temperature rise was found to be quite
reproducible, despite the spread in trigger fields, with a mean
maximum temperature of 4.3 $\pm$ 0.1 K.

In order to compare the time-resolved avalanche measurements with
our earlier SQUID measurements we integrated the calibrated signal
$\dif M/\dif t$ and multiplied it by the sample volume\footnote{The
sample volume $V = 5.5$ mm$^3$ was computed from the saturation
moment $\mu_{\rm S} = 0.77$ mJ\,T$^{-1}$ measured by SQUID
magnetometry, using the known crystal structure and unit cell
volume\cite{Lis-1980} and the known spin $S = 10$ per molecule.} to
obtain the change in magnetic moment $\Delta\mu$.  This yielded
$\Delta\mu = 0.03$ mJ\,T$^{-1}$ for all avalanches, regardless of
the trigger field, which is a great deal smaller than the change in
moment $\Delta\mu\approx 1.5$ mJ\,T$^{-1}$ observed in the SQUID
measurements on the weakly coupled sample.  We believe the
discrepancy to be due to the difference in timescale of the two
measurements. As can be seen in Fig.\ \ref{fig:Avalanche}(a), the
spin reversal rate declines rapidly as the temperature rises,
resulting in the termination of the avalanche after a few
milliseconds. However, the SQUID measurements on the weakly coupled
sample show that spin reversal continues over a much longer period
of $\sim 100$ s at a rate $\dif M/\dif t < 2$
kJ\,T$^{-1}$\,m$^{-3}$\,s$^{-1}$ (close to the noise floor of the
coil).  This suggests that the majority of the spin reversal
detected by the SQUID was due to thermal activation.

Besides avalanches, we also observed slower changes in the
temperature and magnetization at $B=0$ [Fig.\
\ref{fig:Avalanche}(b)]. The change in magnetization at $B=0$ is
known to arise from thermally activated relaxation of a minority
molecular species constituting about 5\% of the
sample.\cite{Evangelisti-2000} The relaxation rate of this species
is much faster than that of the majority species because the
anisotropy barrier is lower. However, the rate of change of the
total magnetic moment is still slow compared to an avalanche,
because thermal activation is a non-cooperative process.  Over a
period of 20 s, the change in magnetic moment was found to be
$\Delta\mu = 0.06$ mJ\,T$^{-1}$, in good agreement with the SQUID
measurements over a similar timescale.  The temperature rise was
found to be very reproducible, with a mean maximum temperature of
2.2 $\pm$ 0.05 K.

In addition to the above observations, we also observed occasional
small signals before or after the main avalanche.  Small temperature
rises of $\sim 50$ mK were observed at $B \approx B_1$ [insets in
Figs.\ \ref{fig:Avalanche}(a) and \ref{fig:Avalanche}(b)] and small
changes in magnetization were observed close to $B \approx B_3$ and
$B_4$. Because these signals were observed only occasionally, we
were unable to synchronize temperature and magnetization
measurements. Therefore, we cannot tell whether they were due to MQT
or miniature avalanches involving parts of the sample in poor
thermal contact with the rest. However, the duration of the
magnetization signals ($\sim 100$ ms) suggests that they are due to
MQT rather than miniature avalanches, which in turn suggests that
the main avalanche does not always result in full spin reversal.

\begin{figure}[!ht]
\begin{center}
\includegraphics{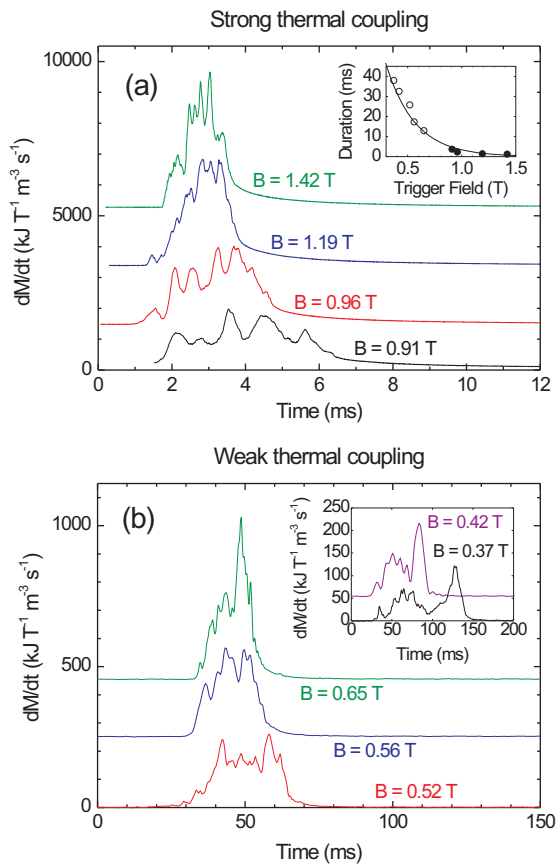}
\caption{(Color online). (a) Avalanches in sample with helium
exchange gas (offset vertically for clarity). Inset: avalanche
duration vs trigger field. Open circles: sample with exchange gas;
closed circles: evacuated sample; solid line: exponential fit. (b)
Avalanches in evacuated sample. Inset: two bursts of spin reversal
are seen at the lowest trigger fields.} \label{fig:HeliumAvalanches}
\end{center}
\end{figure}

We now report the results of a second time-resolved experiment, this
time with the sample enclosed in a glass tube filled with helium
exchange gas to improve the thermal coupling.  The experimental
conditions were identical to those of the previous experiment,
except that we were unable to measure the sample temperature, due to
the difficulty of sealing the glass tube around the sensor leads.
Avalanches were observed at a variety of trigger fields between 0.91
and 1.42 T [Fig.\ \ref{fig:HeliumAvalanches}(a)]. These correspond
approximately to $B_2$ and $B_3$.  The trigger field was
consistently higher than when the glass tube was evacuated [Fig.\
\ref{fig:HeliumAvalanches}(b)], in agreement with our SQUID
measurements. However, the change in magnetic moment $\Delta\mu$ was
similar in both time-resolved experiments, suggesting that, in both
cases, avalanches terminate long before full spin reversal is
achieved.  This is supported by the observation of two closely
spaced bursts of spin reversal at the lowest trigger fields in the
evacuated sample [inset in Fig.\ \ref{fig:HeliumAvalanches}(b)].

We determined the avalanche duration by fitting a Gaussian profile
and taking the full width at half maximum.\footnote{In the case of
the double peaks observed at the lowest trigger fields, only the
first peak was fitted.}  The duration was found to decrease
exponentially as the trigger field increased, independent of thermal
coupling to the bath [inset in Fig.\ \ref{fig:HeliumAvalanches}(a)].
This is consistent with the recently proposed model of magnetic
deflagration. \cite{Suzuki-PRL-2005, Hernandez-Minguez-PRL-2005}  A
particularly interesting observation is that, in the evacuated
sample, avalanches triggered at identical fields had identical
spin-reversal patterns.  Since the lack of smoothness is probably
due to sample inhomogeneity resulting from the layers of varnish
that bind the assembly, this suggests that avalanches propagate in a
very reproducible manner.

In the final part of our report, we investigate the cause of the
temperature rise at $B = 0$.  The amount of heat generated is given
by $\Delta Q = mc\Delta T$, where $m$ is the mass of the sample, $c$
is the specific heat capacity and $\Delta T$ is the observed
temperature rise.  Fominaya et al.\cite{Fominaya-1999} have measured
the specific heat capacity of Mn$_{12}$ as a function of temperature
at zero magnetic field. Using a polynomial fit to extrapolate their
data to temperatures below 3 K we obtain $c = 0.8$
J\,kg$^{-1}$\,K$^{-1}$ at $T = 1.6$ K (the initial temperature) and
$c = 1.3$ J\,kg$^{-1}$\,K$^{-1}$ at $T = 2.2$ K (the peak
temperature). Taking the average of these, we obtain $\langle
c\rangle \approx 1.1$ J\,kg$^{-1}$\,K$^{-1}$. The observed
temperature rise was $\Delta T = 0.6$ K and the sample mass was $m =
17$ mg, leading us to estimate $\Delta Q \approx 11$ $\mu$J.

At $B=0$, no work is done by the external magnetic field because
$\int B\,\dif\mu = 0$. However, work can be done by the internal
demagnetizing field $B_{\rm d} = -\mu_0 N \mu/V$, where $N$ is the
demagnetizing factor, $\mu_0$ is the permeability of free space and
$V$ is the volume of the sample. The magnetostatic energy $E_{\rm m}
= \int B_{\rm d}\,\dif\mu$ is given by
\begin{equation}
E_{\rm m}  = \frac{\mu_0 N \mu^2}{2V}.
\end{equation}
Taking the sample to be approximately spherical we assume a
demagnetizing factor $N = 1/3$.  The total change in the magnetic
moment between full saturation and the beginning of the avalanche is
$\Delta\mu \approx 0.2$ mJ\,T$^{-1}$.  The amount of magnetostatic
energy released is therefore $\Delta E_{\rm m} \approx 10$ $\mu$J,
in good agreement with $\Delta Q$.

In conclusion, we have compared avalanche measurements in samples
with weak and strong thermal coupling to a heat bath and have shown
that a decrease in thermal coupling leads to a reduction in trigger
field. This supports the view that heating is an important factor in
the triggering of avalanches.  Time-resolved measurements of the
sample magnetization have shown an exponential dependence of the
avalanche duration on the trigger field, consistent with the model
of magnetic deflagration. A comparison of time-resolved measurements
with SQUID magnetometry measurements has shown that avalanches
terminate long before saturation is reached.  The elevated sample
temperature following an avalanche allows spin reversal to continue
via thermal activation.

In addition to our studies of spin avalanches, we have also observed
significant heating at $B = 0$.  This has been observed previously
by other researchers,\cite{Suzuki-JAP-2005, Fominaya-1997,
Fominaya-1999} but no satisfactory explanation has been given.  We
attribute the temperature rise to a release of magnetostatic energy.

This work was funded by the UK Department of Trade and Industry
Quantum Metrology Programme, project No.\ QM04.3.4.  The authors
thank the research group of J. Tejada for providing the sample. One
of the authors (A.H.-M.) thanks the Spanish Ministerio de
Educaci\'on y Ciencia for a research grant.

\bibliography{Mn12}

\end{document}